%% file: main.tex
\documentclass[10pt,conference,letterpaper]{IEEEtran} 
\IEEEoverridecommandlockouts
\usepackage{amsmath,amssymb,amsfonts}
\usepackage{algorithmic}
\usepackage{graphicx}
\usepackage{textcomp}
\usepackage{svg}
\usepackage{url}
\usepackage{colortbl}
\usepackage{hhline}
\usepackage{relsize}
\usepackage[hidelinks,colorlinks=true,linkcolor=blue,citecolor=blue]{hyperref}
\usepackage{pifont}

\usepackage{balance}
\usepackage{cite}

\usepackage{hhline}
\usepackage{arydshln}
\usepackage{textcomp}

\usepackage{xcolor}
\usepackage{ifthen}
\usepackage{tikz}
\usepackage{pgfplots,pgfplotstable}
\pgfplotsset{
        set layers={
            background,
            main,
        },
    }
\newlength{\figheight}
\usepackage[square,sort,comma,numbers]{natbib}
\usepackage[T1]{fontenc}

\usetikzlibrary{calc, patterns, spy}

\usepackage{listings}

\usepackage{caption}  
\usepackage{subcaption} 

\usepackage[binary-units=true]{siunitx} 
\usepackage{minted}
\usepackage{listings}
\setminted{fontsize=\small,baselinestretch=1}
\usepackage{booktabs}
\setcitestyle{square}
\usepackage{multirow}
\usepackage{ragged2e}
\usepackage{pifont}
\usepackage{courier}
\usepackage{comment}

\usepackage{circledsteps}
\newcommand\myCircled[2][]{\ifmmode
\Circled[fill color=black,inner color=white,#1]{\mathsf{#2}}
\else
\Circled[fill color=black,inner color=white,#1]{\sffamily#2}
\fi
}

\usepackage{pifont}

\usepackage[hidelinks,colorlinks=true,linkcolor=blue,citecolor=blue]{hyperref}

\usepackage[ruled, lined, linesnumbered, commentsnumbered, longend]{algorithm2e}

\include{acronyms}

\usepackage{svg}

\usepackage{amsmath}
\usepackage{nicematrix}
\input{colors.tex}
\input{comments}

\begin{document}

\title{Efficient In-Memory Acceleration of Sparse Block Diagonal LLMs}

\author{\IEEEauthorblockN{João Paulo C. de Lima\textsuperscript{1,2}, Marc Dietrich\textsuperscript{1}, Jeronimo Castrillon\textsuperscript{1,2}, and Asif Ali Khan\textsuperscript{1}}
\IEEEauthorblockA{\textsuperscript{1}Chair for Compiler Construction, TU Dresden, Germany\\
 \textsuperscript{2}Center for Scalable Data Analytics and Artificial Intelligence (ScaDS.AI), Dresden, Germany \\
Corresponding authors: \{joao.lima, asif\_ali.khan\}@tu-dresden.de}
\vspace{-1.5em}
}


\maketitle

\begin{abstract}
Structured sparsity enables deploying large language models (LLMs) on resource-constrained systems. Approaches like dense-to-sparse fine-tuning are particularly compelling, achieving remarkable structured sparsity by reducing the model size by over 6.7$\times$, while still maintaining acceptable accuracy. Despite this reduction, LLM inference, especially the decode stage being inherently memory-bound, is extremely expensive on conventional Von-Neumann architectures. Compute-in-memory (CIM) architectures mitigate this by performing computations directly in memory, and when paired with sparse LLMs, enable storing and computing the entire model in memory -- eliminating the data movement on the off-chip bus and improving efficiency. 
Nonetheless, naively mapping sparse matrices onto CIM arrays leads to poor array utilization and diminished computational efficiency. In this paper, we present an automated framework with novel mapping and scheduling strategies to accelerate sparse LLM inference on CIM accelerators. By exploiting block-diagonal sparsity, our approach improves CIM array utilization by over 50\%, achieving more than $4\times$ reduction in both memory footprint and the number of required floating-point operations.
\end{abstract}


\input{content/introduction}
\input{content/background}

\input{content/mapping}

\input{content/results}

\input{content/related-work}
\input{content/conclusions}

\bibliographystyle{IEEEtran}
\bibliography{main}

\end{document}

%% file: colors.tex
\definecolor{pairedOneLightBlue}{HTML}{a6cee3}
\definecolor{pairedTwoDarkBlue}{HTML}{1f78b4}
\definecolor{pairedThreeLightGreen}{HTML}{b2df8a}
\definecolor{pairedFourDarkGreen}{HTML}{33a02c}
\definecolor{pairedFiveLightRed}{HTML}{fb9a99}
\definecolor{pairedSixDarkRed}{HTML}{e31a1c}
\definecolor{butter1}{rgb}{0.988,0.914,0.310}
\definecolor{butter2}{rgb}{0.929,0.831,0.000}
\definecolor{butter3}{rgb}{0.769,0.627,0.000}
\definecolor{orange1}{rgb}{0.988,0.686,0.243}
\definecolor{orange2}{rgb}{0.961,0.475,0.000}
\definecolor{orange3}{rgb}{0.808,0.361,0.000}
\definecolor{chocolate1}{rgb}{0.914,0.725,0.431}
\definecolor{chocolate2}{rgb}{0.757,0.490,0.067}
\definecolor{chocolate3}{rgb}{0.561,0.349,0.008}
\definecolor{chameleon1}{rgb}{0.541,0.886,0.204}
\definecolor{chameleon2}{rgb}{0.451,0.824,0.086}
\definecolor{chameleon3}{rgb}{0.306,0.604,0.024}
\definecolor{skyblue1}{rgb}{0.447,0.624,0.812}
\definecolor{skyblue2}{rgb}{0.204,0.396,0.643}
\definecolor{skyblue3}{rgb}{0.125,0.290,0.529}
\definecolor{plum1}{rgb}{0.678,0.498,0.659}
\definecolor{plum2}{rgb}{0.459,0.314,0.482}
\definecolor{plum3}{rgb}{0.361,0.208,0.400}
\definecolor{scarletred1}{rgb}{0.937,0.161,0.161}
\definecolor{scarletred2}{rgb}{0.800,0.000,0.000}
\definecolor{scarletred3}{rgb}{0.643,0.000,0.000}
\definecolor{aluminium1}{rgb}{0.933,0.933,0.925}
\definecolor{aluminium2}{rgb}{0.827,0.843,0.812}
\definecolor{aluminium3}{rgb}{0.729,0.741,0.714}
\definecolor{aluminium4}{rgb}{0.533,0.541,0.522}
\definecolor{aluminium5}{rgb}{0.333,0.341,0.325}
\definecolor{aluminium6}{rgb}{0.180,0.204,0.212}

\definecolor{blind_safe_one_scheme_three_colors}{RGB}{102,194,165}
\definecolor{blind_safe_two_scheme_three_colors}{RGB}{252,141,98}
\definecolor{blind_safe_three_scheme_three_colors}{RGB}{141,160,203}

\definecolor{blind_safe_one_scheme_four_colors}{RGB}{166,206,227}
\definecolor{blind_safe_two_scheme_four_colors}{RGB}{31,120,180}
\definecolor{blind_safe_three_scheme_four_colors}{RGB}{178,223,138}
\definecolor{blind_safe_four_scheme_four_colors}{RGB}{51,160,44}

\definecolor{blind_safe_one_scheme_five_colors}{RGB}{240,249,232}
\definecolor{blind_safe_two_scheme_five_colors}{RGB}{186,228,188}
\definecolor{blind_safe_three_scheme_five_colors}{RGB}{123,204,196}
\definecolor{blind_safe_four_scheme_five_colors}{RGB}{67,162,202}
\definecolor{blind_safe_five_scheme_five_colors}{RGB}{8,104,172} 

\definecolor{blind_safe_one_scheme_seven_colors_grnblu}{RGB}{240,249,232}
\definecolor{blind_safe_two_scheme_seven_colors_grnblu}{RGB}{204,235,197}
\definecolor{blind_safe_three_scheme_seven_colors_grnblu}{RGB}{168,221,181}
\definecolor{blind_safe_four_scheme_seven_colors_grnblu}{RGB}{123,204,196}
\definecolor{blind_safe_five_scheme_seven_colors_grnblu}{RGB}{78,179,211}
\definecolor{blind_safe_six_scheme_seven_colors_grnblu}{RGB}{43,140,190}
\definecolor{blind_safe_seven_scheme_seven_colors_grnblu}{RGB}{8,88,158}

\definecolor{blind_safe_one_scheme_seven_colors}{RGB}{118,42,131}
\definecolor{blind_safe_two_scheme_seven_colors}{RGB}{175,141,195}
\definecolor{blind_safe_three_scheme_seven_colors}{RGB}{231,212,232}
\definecolor{blind_safe_four_scheme_seven_colors}{RGB}{247,247,247}
\definecolor{blind_safe_five_scheme_seven_colors}{RGB}{217,240,211}
\definecolor{blind_safe_six_scheme_seven_colors}{RGB}{127,191,123}
\definecolor{blind_safe_seven_scheme_seven_colors}{RGB}{27,120,55}

\definecolor{yellow_one}{RGB}{255,255,212}
\definecolor{yellow_two}{RGB}{254,217,142}
\definecolor{yellow_three}{RGB}{254,153,41}
\definecolor{yellow_four}{RGB}{217,95,14}
\definecolor{yellow_five}{RGB}{153,52,4}

\definecolor{minted_red_bg}{HTML}{ffdfde}
\definecolor{minted_blue_bg}{HTML}{dcdff7}

\definecolor{minted_ins_norm}{HTML}{ffffff}
\definecolor{minted_ins_mram}{HTML}{facb89}
\definecolor{minted_ins_wram}{HTML}{f5c9e5}
\definecolor{minted_ins_cf}{HTML}{b4e0ab}
\definecolor{minted_ins_comp}{HTML}{afccfa}

\definecolor{codegray}{rgb}{0.5,0.5,0.5}
\definecolor{lightblue}{HTML}{deebff}

\definecolor{one_scheme_seven_colors}{HTML}{c7e9b4}
\definecolor{two_scheme_seven_colors}{HTML}{7fcdbb}
\definecolor{three_scheme_seven_colors}{HTML}{41b6c4}
\definecolor{four_scheme_seven_colors}{HTML}{1d91c0}
\definecolor{five_scheme_seven_colors}{HTML}{225ea8}
\definecolor{six_scheme_seven_colors}{HTML}{253494}
\definecolor{seven_scheme_seven_colors}{HTML}{081d58}

%% file: comments.tex
\usepackage{xcolor}
\usepackage{ifthen}
\usepackage{mathabx}
\newboolean{showcomments}
\setboolean{showcomments}{false}
\ifthenelse{\boolean{showcomments}}
{ \newcommand{\mynote}[3]{
     \fbox{\bfseries\sffamily\scriptsize#1}   {\small$\blacktriangleright$\textsf{\emph{\color{#3}{#2}}}$\blacktriangleleft$}}}
{ \newcommand{\mynote}[3]{}}
\definecolor{asparagus}{rgb}{0.53, 0.66, 0.42}
\newcommand{\jc}[1]{\mynote{JC}{#1}{asparagus}}

\newcommand{\ak}[1]{\mynote{AK}{#1}{blue}}
\newcommand{\jp}[1]{\mynote{JP}{#1}{orange}}

%% file: content/introduction.tex
\section{Introduction}
\label{sec:intro}

Pre-trained transformer models, with billions of parameters, demand substantial computational and memory resources, making them impractical for resource-constrained devices~\cite{radford2019language, devlin2019bert}. Structured sparsity significantly reduces these requirements, enabling potential deployment on such systems~\cite{tan2024more, dao2022monarch}. Dense-to-sparse fine-tuning transforms pre-trained dense models into sparse models with comparable accuracy, achieving up to $4-8\times$ compression depending on the structure and task~\cite{dao2022monarch, tan2024more}. However, sparsity alone is not sufficient for efficient execution. Hardware-level inefficiencies, especially in exploiting the block-structured sparsity, remain a critical bottleneck to achieving performance gains.

Even with reduced model sizes, transformer inference, particularly memory-bound in the decoding phase, incurs high energy costs due to data movement in conventional von Neumann systems, consuming over 62\% of total energy~\cite{data_movement_energy}. Analog compute-in-memory (CIM) accelerators address this by performing computations directly within memory, leveraging crossbar arrays where memory cells serve as both storage and computational elements~\cite{ahmed2019compiler,de2022quantization,cinm-landscape}. These arrays enable constant-time matrix-vector multiplication (MVM) in-place, exploiting the analog properties of memory devices to perform operations without data transfer. This eliminates the von Neumann bottleneck and drastically reduces energy and latency~\cite{hu2025cross}. 

However, on resource-constrained devices, the large GEMM sizes in transformer models still pose challenges, requiring sparsification to make them fit in limited CIM array capacities. Structured approaches such as Monarch matrices~\cite{dao2022monarch} replace dense weight matrices in components like attention and feedforward layers with structured, sparse representations. Monarch matrices, which are products of two block-diagonal matrices (up to permutation), enable sub-quadratic multiplication and can represent common operations such as FFTs and convolutions~\cite{dao2021pixelated}. Compared to unstructured sparsity, this structured form offers both computational advantages and expressiveness. Nonetheless, mapping these block-diagonal matrices efficiently onto CIM arrays is non-trivial. Naive mappings often lead to severe underutilization of array capacity, resulting in performance and energy inefficiencies. Addressing this challenge requires a careful co-design of sparsity patterns, data layout, and memory-aware mapping strategies to fully realize the benefits of sparse models on CIM accelerators.

To this end, we propose an automated framework that transforms the dense layers of transformer networks into structured sparse representations and efficiently deploys them onto analog CIM designs. Specifically, our contributions are:
\begin{itemize}
    \item \textbf{CIM-aware sparse mapping:} We propose latency- and capacity-optimized mapping strategies to densely pack block-diagonal structured matrices onto CIM arrays. These strategies reduce array fragmentation and enable selective activation of crossbar rows and columns, thus reducing execution time and energy consumption by requiring fewer and lower-precision analog-to-digital converters (ADCs) (Sec.~\ref{subsec:mapping}). 
     \item \textbf{Performance-aware scheduling:} Our scheduling module selectively activates rows within the crossbar arrays, intelligently balancing ADCs sharing and parallelism, leading to significant reductions in energy consumption and inference latency (Sec.~\ref{subsec:scheduler}).
     \item \textbf{End-to-end automation:} Our framework integrates dense-to-sparse (D2S) transformation, crossbar mapping, and scheduling into an automated toolchain, targeting analog CIM accelerators (Fig.~\ref{fig:mm-flow}).
    
\end{itemize}

We conduct design space exploration by analyzing how different mappings requiring varying DA/AD converter precisions and sharing degrees affect performance and energy consumption, and validate our strategies across multiple models. 
Our evaluations, compared to dense models on the CIM baseline, show that our framework reduces execution time and energy consumption by over 1.7$\times$, while also simultaneously reducing the memory footprint by over 4$\times$. 

%% file: content/background.tex
\section{Background} 
\label{sec:bg}
This section provides background on analog CIM, transformer networks, and structured sparsity.  

\subsection{Analog Computing-in-Memory}
\label{subsec:cim-bg}
\begin{figure}
    \centering
    \includegraphics[trim={0 0.8cm 0 0.9cm},clip,width=0.7\columnwidth]{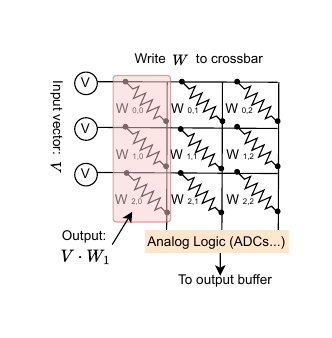}
    \caption{Dot product on an analog CIM crossbar}
    \label{fig:cimarray}
\end{figure}

Analog CIM leverages the physical properties of memory devices to multiply an input vector $V$ with a matrix $W$ directly within memory arrays. This can be implemented using nonvolatile memory (NVM) as well as SRAM technologies. Fig.~\ref{fig:cimarray} shows a typical NVM-based CIM crossbar, where the matrix $W$ is programmed into the memory cells, and the input vector $V$ is applied as voltages across the rows (wordlines). The MVM is performed via Ohm's and Kirchhoff's laws: the accumulated current on each vertical line (bitline) corresponds to the dot product of $V$ with the respective column of $W$ (e.g., column $W_1$ is highlighted in red). 

SRAM-based analog CIM follows similar principles but is implemented in different ways, such as current~\cite{dong202015}, charge-based~\cite{lee2021fully}, or time-based~\cite{wu202228nm}. Regardless of the underlying technology, analog CIM follows a weight-stationary dataflow consisting of the following steps: (1) converting digital inputs into analog signals using digital-to-analog converters (DACs), typically bit-streamed over multiple cycles; (2) performing multiplication in the analog domain ($V \cdot W$);
(3) summing the partial products in the analog domain; (4) converting the analog results back to digital using ADCs; and (5) accumulating outputs from multiple crossbars using shift-and-add circuits.

ADC scalability remains a major bottleneck in analog CIM, often accounting for up to 60\% of energy and nearly 80\% of area in CIM accelerators~\cite{negi2022nax}. Due to their large footprint, ADCs are typically shared across multiple columns, which limits the overall throughput of the crossbar. Successive Approximation Register (SAR) ADCs are commonly used in CIM designs; they allow precision to be dynamically adjusted by changing the number of comparison steps during conversion~\cite{nag2018newton}. Reducing the number of conversion steps lowers resolution and power consumption. By intelligently leveraging model compression techniques, especially block-diagonal sparsity patterns, it becomes possible to aggressively share ADCs across inactive columns and reduce the required ADC precision. Lower ADC precision directly translates to reduced energy consumption and smaller area footprint, thereby improving energy and area efficiency of CIM accelerators~\cite{adc-precision}.

\subsection{Transformer Networks}
\label{subsec:transformer-bg}
Transformer models are built upon the self-attention mechanism, which enables them to capture long-range dependencies and contextual relationships between tokens (e.g., words) in a sequence (e.g., a sentence). At the heart of this mechanism is the scaled dot-product attention (SPDA), which operates on three sets of vectors: query ($q$), keys ($\mathbf{K}$), and values ($\mathbf{V}$). The attention output is a weighted sum of the value vectors, where the weights are determined by the similarity between the query and each key vector. This similarity is computed as the dot product between $q$ and each key $k_j \in \mathbf{K}$, scaled by the square root of the key dimensionality $d_k$:
\[
\text{Attention}(q, \mathbf{K}, \mathbf{V}) = \text{softmax}\left( \frac{q \mathbf{K}^\top}{\sqrt{d_k}} \right) \mathbf{V}.
\]

To improve the model's ability to attend to information from different representation subspaces, Transformer architectures employ multi-head attention (MHA). Instead of performing a single attention operation, MHA projects the input queries, keys, and values into multiple lower-dimensional subspaces using learnable projection matrices $\mathbf{W}^Q_i$, $\mathbf{W}^K_i$, and $\mathbf{W}^V_i$ for each attention head $i$. The SDPA operation is then applied independently in each head, allowing the model to jointly attend to information from different perspectives. The outputs from all heads are subsequently concatenated and linearly transformed to form the final output~\cite{vaswani2017attention}.  Following the attention mechanism, each Transformer layer includes a feed-forward network (FFN) composed of two linear transformations with a nonlinear activation in between.

\begin{figure*}[tbh]
    \centering
    \includegraphics[width=\textwidth]{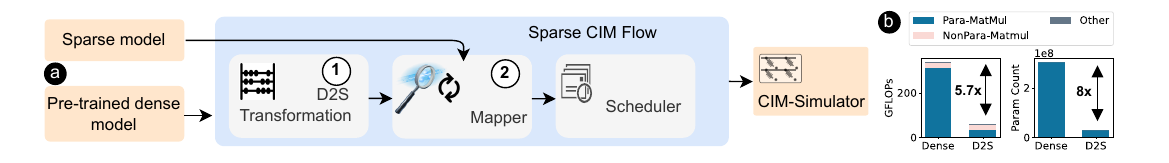}
    \caption{(a) Overview of the proposed framework; (b) the expected reduction in FLOPs and parameter count for BERT-large.}
    \label{fig:mm-flow}
\end{figure*}

\subsection{Monarch Factorization}
\label{subsec:monarch-factor}
Structured matrix factorizations have been well studied for D2S transformations, with Butterfly factorizations~\cite{dao2019learning,li2015butterfly} offering compact representations of linear operators through hierarchical and diagonal sparsity. These enable efficient MVM and approximation with reduced time and space complexity. Building upon this foundation, the Monarch framework~\cite{dao2022monarch} introduces a structured matrix class that expresses dense weight matrices as a product of two block-diagonal matrices interleaved with fixed permutations. This structure is highly expressive, hardware-efficient, and capable of approximating dense matrices with minimal loss in representational fidelity.

In its most general form, an order-$p$ Monarch matrix is structured as a product $M = \prod_{i=1}^p (P_i B_i) P_0$, where $M \in \mathbb{R}^{n\times n}$, $P_i$ are  alternate permutations, and $B_i$ are block-diagonal matrices~\cite{fu2023monarch}. In practice, existing works have focused on factorizations with two block-diagonal matrices (i.e., p=2) composed of blocks of size $b \times b$, where $b = \sqrt n$~\cite{dao2022monarch,fu2023monarch}. 
This yields a subquadratic matrix multiplication complexity of $\mathcal{O}(pn^{(p+1)/p})$, which is highly expressive for common transforms (e.g., it generalizes FFT) and sufficiently expressive for attention and FFN layers. 

In addition to algorithmic efficiency, Monarch matrices are hardware-friendly: their structure decomposes larger GEMMs into smaller GEMMs, which lend themselves well to tensor cores in modern GPUs. This paper explores their execution on CIM accelerators, where the fine-grained block structure of Monarch matrices is leveraged for improved mapping, parallelism, and energy efficiency.

%% file: content/mapping.tex
\section{Monarch Matrices Acceleration Using CIM}
\label{sec:mapMonarch}
Fig.~\ref{fig:mm-flow}a illustrates the overall framework: starting from a pretrained dense model, the dense-to-sparse (D2S) transformation generates structured sparse matrices, which are then \emph{mapped} onto CIM arrays. The \textit{scheduler} orchestrates the execution by issuing memory commands that balance ADC sharing and parallelism, optimizing the trade-off between energy and latency. For models that have already been trained or fine-tuned with Monarch sparsity, the D2S step is not required. On BERT-large with 512-token input sequences, D2S reduces parameters by $8\times$ and FLOPs by 5.7$\times$ compared to the baseline (\emph{Dense}), as illustrated in Fig.~\ref{fig:mm-flow}b. The following sections provide detailed explanations of each step in the framework.

\begin{figure}[h]
    \centering

    \includegraphics[width=1.05\columnwidth]{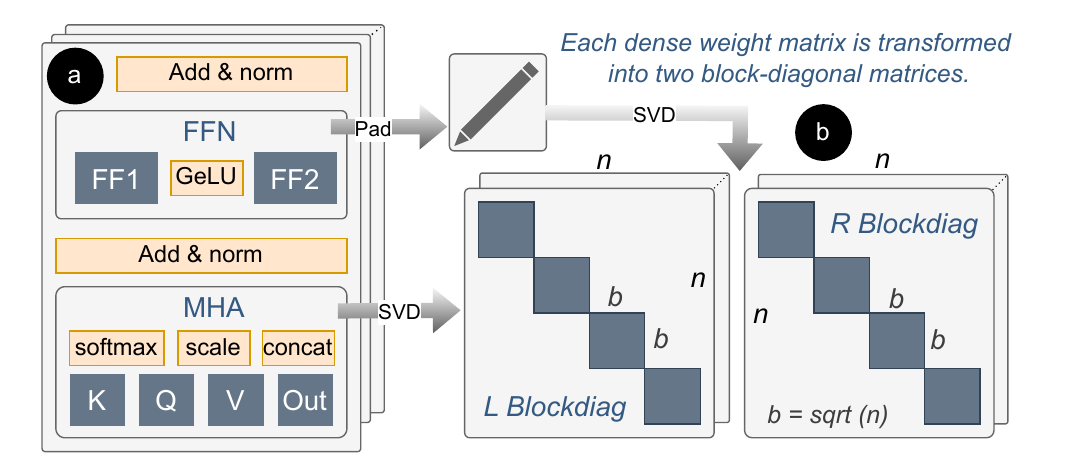}

    \caption{D2S transformation, (a) dense model (b) sparse.}
    \label{fig:d2s}
\end{figure}

\subsection{Dense-to-sparse (D2S) Transformation}
\label{subsec:d2stransform}


To enable D2S transformation without retraining, authors in~\cite{dao2022monarch} propose an analytical method to approximate a dense matrix \( W \in \mathbb{R}^{n \times n} \) with a Monarch-structured matrix \( M \in \mathbb{R}^{n \times n} \) by solving an optimization problem that minimizes the Frobenius norm \( \|W - M\|_F \). This is achieved by reshaping \( W \) into slices that match the Monarch block structure and applying rank-1 singular value decomposition (SVD) to obtain the factor matrices. The resulting decomposition, 

\begin{equation}\label{eq:monarchMatrix}
M = P \cdot L \cdot P \cdot R \cdot P,
\end{equation}

use block-diagonal matrices $L$ and $R$ derived from SVD, and a fixed permutation matrix $P$. This structured approximation preserves the accuracy of the original matrix while significantly improving memory and compute efficiency. An overview of the factorized matrices $L$ and $R$ is shown in Fig.~\ref{fig:d2s}.

As shown, we apply the D2S transformation only to parameterized matmuls, such as the linear projections in the MHA and FFN blocks (\emph{Para-Matmul} in Fig.~\ref{fig:mm-flow}b). In contrast, non-parameterized matmul (\emph{NonPara-Matmul}), such as those used to compute attention scores and apply attention weights, operate solely on activation tensors and remain untransformed.

\subsection{Mapping Sparse Matrices onto CIM Arrays}
\label{subsec:mapping}

As discussed in Sec.~\ref{subsec:cim-bg}, MVM in CIM typically involves pre-storing the weight matrix \( W \) within the memory array, while the input vector \( X \) is encoded as input voltage and applied to all array rows. After applying the D2S transformation described in Sec.~\ref{subsec:d2stransform}, the MVM operation transforms from \( X \cdot W \) to \( X \cdot M \), where \( M \) is a Monarch-structured matrix defined in Eq.~\ref{eq:monarchMatrix}. This requires mapping the resulting block-diagonal sparse matrices ($L, R$) onto CIM arrays.

Let the CIM array have dimension \( m \times m \), and assume that each block in the \( L \) and \( R \) matrices is of size \( b = \sqrt{n} \). Consider a transformer model with \( N \) layers, each containing \( p \) parameterized matrix multiplications. The mapping problem then reduces to placing \( N \cdot p \) Monarch matrices onto the available CIM arrays of size \( m \times m \). 
\jc{This sounds like somethign that could be formalized as ILP or alike with different possible min/max targets (for the extension as journal $;-)$)}

In the following, we present two mapping strategies: one optimized for performance, and another optimized for crossbar utilization and energy efficiency. For simplicity, we demonstrate and describe the mapping of a single matrix, though this applies to all matrices. 

\subsubsection{Performance Optimized Mapping}
\label{subsubsec:naive}
\begin{figure}[h]
    \centering

    \includegraphics[width=1.02\columnwidth]{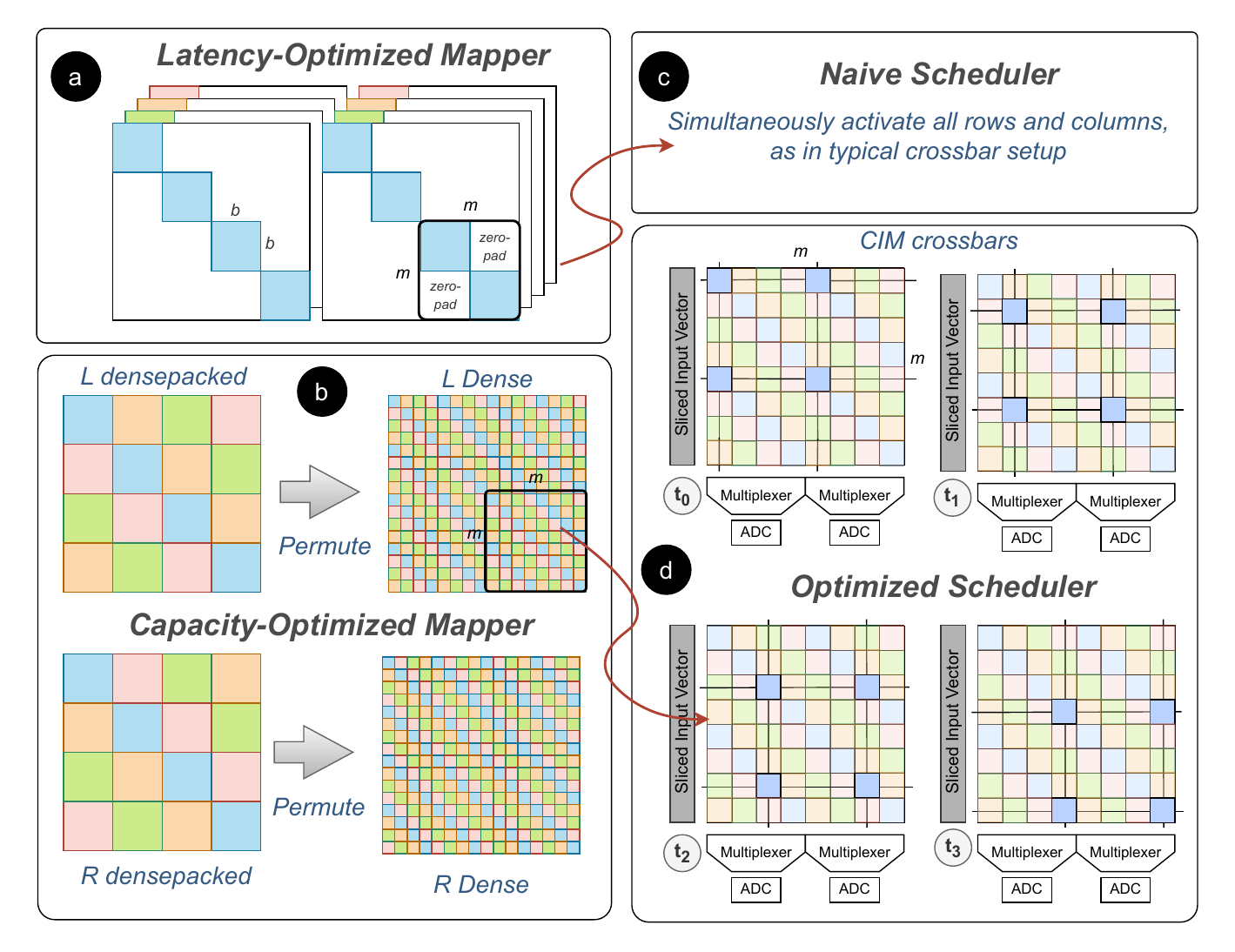}

    \caption{Proposed mapping and scheduling strategies.}
    \label{fig:mapping-sched}
\end{figure}

In this approach, we \emph{naively} map each block-diagonal matrix to one or more CIM arrays without considering the memory utilization. Given Monarch matrices with block size \( b \) and CIM arrays of size \( m \times m \), the \( L \) and \( R \) matrices are partitioned based on \( m \), and each partition, comprising one or more blocks, is assigned to a separate CIM array. If \( b = m \), this results in a one-to-one mapping, fully utilizing the array capacity. However, in practice, \( b \ll m \), which allows packing multiple blocks per array, but still leads to significant underutilization. For instance, in Fig.~\ref{fig:mapping-sched}a, when \( m = 2b \), two blocks can be packed into a single array, with remaining cells zero-padded. This enables parallel execution of blocks since the padded zeros do not impact correctness. Nonetheless, it comes at the cost of memory underutilization, which is especially a critical concern in resource-constrained systems.

In general, the total number of CIM arrays required to store the \( L \) and \( R \) matrices is approximately \( n / m \). The effective utilization of each array, defined as the ratio of non-zero (valid) entries to total array capacity, is given by \( (b/m) \times 100\)\%. For example, with \( n = 1024 \), \( m = 256 \), and \( b = 32 \), the effective array utilization is only \( (32 / 256) \times 100\% = 12.5\% \), meaning that 87.5\% of the array remains unused due to zero-padding. 

This mapping also implicitly assumes that the number of CIM arrays in the system is larger than the required number of arrays, which may not be the case in resource-constrained systems. For systems with a limited number of CIM arrays, this mapping requires rewriting array data (swapping it with new data) dynamically during execution, which incurs significant overhead, especially in NVM-based CIM systems. We next introduce a capacity-optimized mapping strategy that densely packs the sparse block-diagonal matrices in CIM arrays to maximize array utilization and minimize write overhead.

\subsubsection{Capacity Optimized Mapping}
\label{subsubsec:opt-mapping}

This mapping strategy leverages the structured sparsity inherent in Monarch matrices to densely pack multiple block-diagonals into a single CIM array. Given that Monarch-structured matrices reduce parameter count by a factor of \( \sqrt{n} \), it is possible to store up to \( \sqrt{n} \) block-diagonal matrices in a single \( m \times m \) CIM array. As illustrated in Fig.~\ref{fig:mapping-sched}b, the densely packed arrays for \( L \) and \( R \) accommodate four block-diagonals each. These diagonals may correspond to different parameterized operations within a transformer layer or to partitions of a single large matrix that has been partitioned to match array dimensions. This maximizes array utilization, approaching 100\% when \( m \) is a multiple of \( b \), and substantially reduces the number of CIM arrays needed for in-place inference. 

\begin{figure}[h]
    \centering

    \begin{subfigure}[b]{0.28\columnwidth}
        \centering
        \includegraphics[width=\columnwidth]{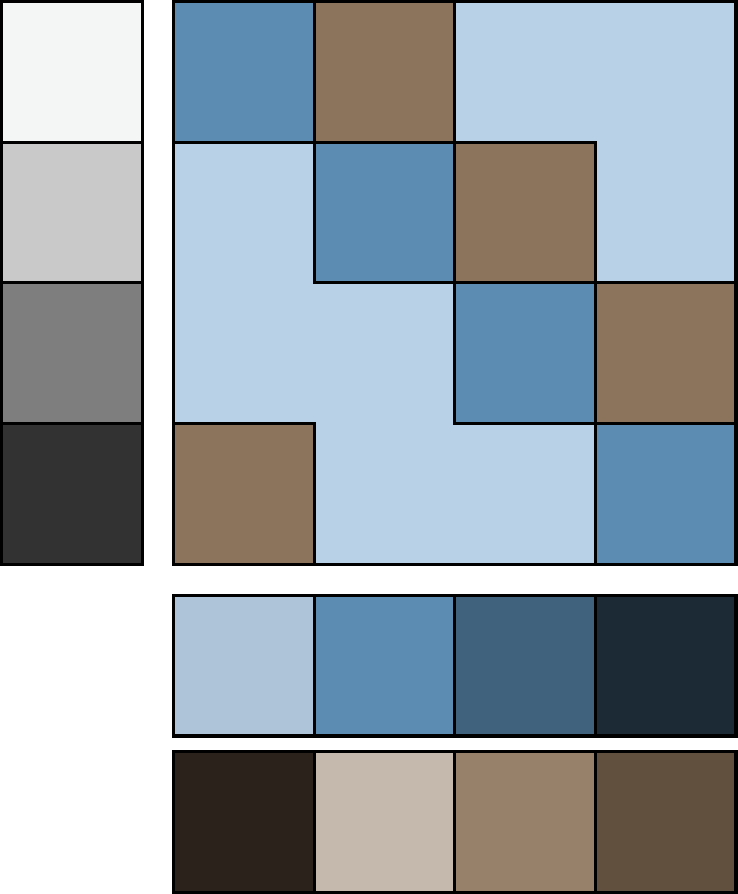}
        \caption{Diagonal rotation}
        \label{fig:rotation}
    \end{subfigure}
    \hfill 
    \begin{subfigure}[b]{0.33\columnwidth}
        \centering
        \includegraphics[width=\columnwidth]{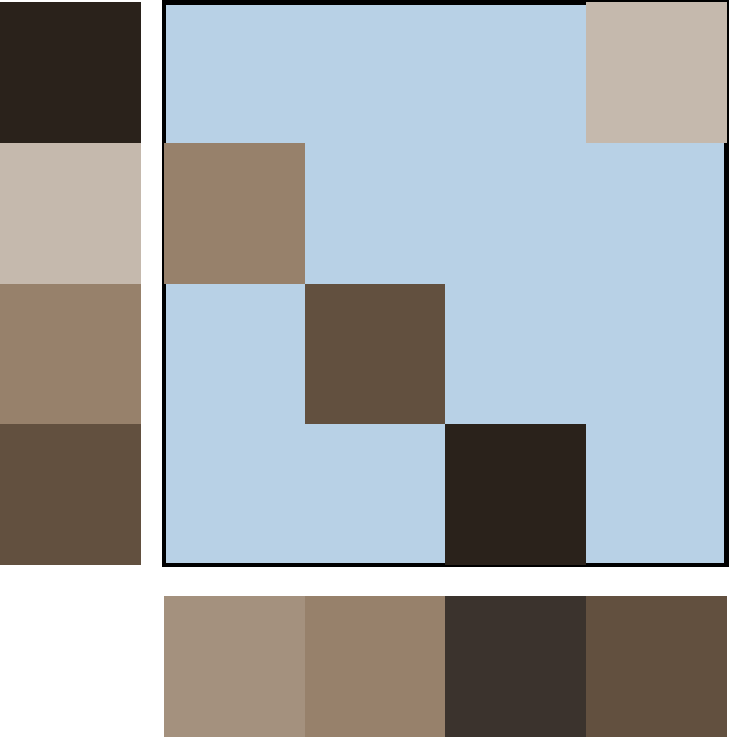}
        \caption{No shifting (wrong)}
        \label{fig:no_shifting}
    \end{subfigure}
    \hfill 
    \begin{subfigure}[b]{0.33\columnwidth}
        \centering
        \includegraphics[width=\columnwidth]{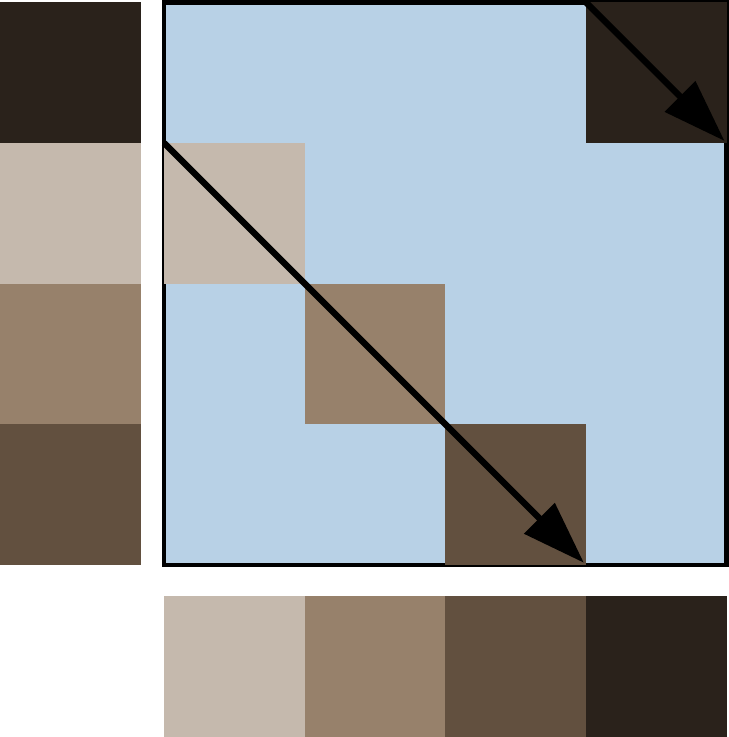}
        \caption{Correct shifting}
        \label{fig:corr_shifting}
    \end{subfigure}

    \caption{An example to show why rotations (a) and shifting are needed (b, c)}
    \label{fig:shifting}
\end{figure}

\paragraph{Handling Rotations and Shifts}
Each block-diagonal in our densely packed CIM array is assigned a unique diagonal index \( i \), indicating its relative position within the array. During the MVM operation, this indexing introduces block-wise cyclic rotations in the output vector. Specifically, a diagonal at index \( i \) produces an output that is rotated by \( i \) positions. For instance, in Fig.~\ref{fig:rotation}, the input vector (represented in gray) multiplied by the block-diagonal at index 0 (blue color) produces results that do not require any rotation. However, the same vector applied to the diagonal at index 1 (brown color) produces results that need to be rotated left by one to produce the correct final result.

Given the two-stage structure of Monarch matrices, first multiplying with \( L \), then with \( R \), it is possible to exploit rotational symmetry to cancel out these rotations. By carefully choosing the diagonal index \( i_R \) for each block in \( R \) such that:
\[
i_R = -i_L \mod b,
\]
the rotation introduced by the first stage (\( L \)) is effectively neutralized in the second stage (\( R \)). This reduces the need for costly post-processing or rotation correction operations and enables more efficient scheduling on the CIM hardware.

To exploit this symmetry, the blocks in the \( R \) stage need to be \emph{shifted} to align with the rotational offsets from the \( L \) stage. As shown in Fig.~\ref{fig:corr_shifting}, the output blocks of \( L \) must be routed or reassigned to the appropriate diagonals in \( R \) based on the rotation index \( i_L \). This reassignment ensures that the composition \( R \cdot L \) yields a correctly aligned output vector.

Special care is required for diagonal indices \( i = 0 \) and \( i = b/2 \), as these indices are self-inverses under modulo \( b \) and thus cannot be paired within the same dense matrix without violating the symmetry constraint. These special cases must be distributed across different Monarch matrices or carefully managed to avoid redundant or conflicting assignments.

\subsubsection{Folding Permutations into the Matrix Structure}
\label{ss:permutation}
Recall from Sec.~\ref{subsec:d2stransform} that \( M = P \cdot L \cdot P \cdot R \cdot P \), i.e., in addition to the $L, R$ multiplication, the permutation matrix $P$ must also be applied three times in correct order to produced the correct result. However, these inherent permutations in the Monarch structure can be embedded directly into the matrix structure itself to eliminate the need for explicit permutation operation. The conventional monarch matrix requires three separate permutation operations surrounding the sparse block-diagonal matrices. However, by algebraic rearrangement, this can be rewritten as: 
\[
M = (P L P) \cdot P \cdot (P R P),
\]
which folds the outer permutations directly into the \( L \) and \( R \) matrices. This transformation reduces the number of permutation steps from three to just one, and can be performed offline before mapping the $L, R$ matrices to CIM arrays. 

More importantly, CIM arrays typically have fewer ADCs than the number of bitlines, i.e., ADCs are shared across groups of bitlines, and an input vector is multiplied in the CIM array over multiple cycles. The transformed $M$ structure naturally aligns with this multi-cycle behavior by enabling multiplexed ADC access, thereby improving throughput. 

\subsection{The Scheduling Module}
\label{subsec:scheduler}
For the latency-optimized mapping (Sec.~\ref{subsubsec:naive}), all blocks can be computed in parallel by simultaneously activating all rows and columns of the CIM array, as in a typical crossbar operation. In contrast, the capacity-optimized mapping strategy presented in Sec.~\ref{subsubsec:opt-mapping} requires the scheduler to issue mapping-aware CIM instructions to ensure correct execution. Naively, activating all rows and columns in this mapping would lead to incorrect results, as each column stores data from multiple block-diagonal matrices.

To illustrate this, consider the highlighted $m \times m$ partition (where $m=8$) of the $L$ matrix in Fig.~\ref{fig:mapping-sched}b, which is mapped to a CIM array as shown in Fig.~\ref{fig:mapping-sched}d. Note that the figure shows the same array state at four different timestamps. Based on the diagonal indices of the block-diagonal matrix, the computations are temporally scheduled. At time \textcircled{\scriptsize $t_0$}, only two rows and two columns are activated, corresponding to four active cells (highlighted in blue). At \textcircled{\scriptsize $t_1$}, the next four cells along the same diagonal—shifted down by one position—are computed. This pattern continues so that the subsequent elements along the diagonal are processed at \textcircled{\scriptsize $t_2$} and \textcircled{\scriptsize $t_3$}, completing the computation for one diagonal. The same procedure is repeated for the remaining diagonals in the array. 

The scheduler, aware of the underlying memory mapping and block-diagonal sparsity, automatically generates the correct memory addresses and issues the appropriate control commands to execute the matrix multiplication accurately. It is important to note that, while computations within a single CIM array are performed sequentially due to temporal scheduling, all CIM arrays operate in parallel. Since the array dimensions are typically much smaller than the matrix dimensions, each array produces partial results that are subsequently combined to compute the final output.

%% file: content/results.tex
\section{Experimental Setup and Evaluation}
\label{sec:eval}
This section describes our experimental setup, benchmarks, and a discussion of the results. 
~\\\noindent\textbf{Benchmarks:} To evaluate our proposed framework, we use the parameterized matmults in three different transformer models: BERT-large~\cite{devlin2019bert}, BART-large~\cite{lewis2019bart}, and GPT-2-Medium~\cite{radford2019language}. BART is an encoder-decoder model, while GPT-2 and BERT are decoder-only and encoder-only transformer architectures, respectively. The context lengths are 512 for BERT, and 1024 for both BART and GPT-2.

\noindent\textbf{Simulated system:} We use the open-source simulator from~\cite{buchel2025efficient}, which models 2D and 3D analog in-memory computing (AIMC) accelerators with configurable architectural parameters and execution flows. The simulator supports transformer model deployment across weight-stationary CIM arrays for MVM operations, digital processing units (DPUs) for auxiliary operations, and dedicated multi-head attention (MHA) units. It faithfully models data movement and scheduling costs, and enables energy and latency estimation. Since the monarch factorization only targets parameterized matmuls, which constitute over 80\% of FLOPs (see Fig.~\ref{fig:mm-flow}b), we specifically focus on the performance of parameterized ones.  

The underlying CIM technology used in the simulator is based on IBM's PCM, with configuration details listed in Table~\ref{tab:params}. However, our proposed mapping and scheduling strategies are CIM technology-agnostic and will deliver similar benefits on other technologies as well. We also compare our results to the NVIDIA RTX 3090 Ti GPU. \ak{to double check!}\jp{For space reasons, I'd just mention the model}

\begin{table}[htb]

\caption{Baseline CIM parameters for $d_{model}$=1024}
\begin{center}
{\small
\begin{tabular}{c|r|r} 
\toprule
Specification & Latency (\SI{}{\nano\second}) & Energy (\SI{}{\nano\joule}) \\
\midrule
MVM (256$\times$256 PCM) & 100 & 10 \\
ADCs SAR (8b)~\cite{shafiee2016isaac} & 0.833 &  13.33e$^{-3}$ \\
Communication & 48 & 51.7 \\
LayerNorm & 100 & 42 \\
ReLU / GeLU / Add & 1 / 70 / 36 & 0.06 / 38.5 / 37.7 \\
\bottomrule
\end{tabular}
}
\label{tab:params}
\end{center}
\end{table}

\noindent\textbf{Mapping \& scheduling strategies:} We evaluate three setups: 
\begin{itemize}
   \item \textit{Linear:} Maps the linear layers of the dense pre-trained model onto the CIM accelerator and serves as the baseline for the proposed mapping strategies.
    \item \textit{SparseMap:} A latency-optimized sparse mapping, as described in Sec.~\ref{subsubsec:naive}.
    \item \textit{DenseMap:} A capacity-optimized mapping strategy, discussed in Sec.~\ref{subsubsec:opt-mapping}.
     
\end{itemize}

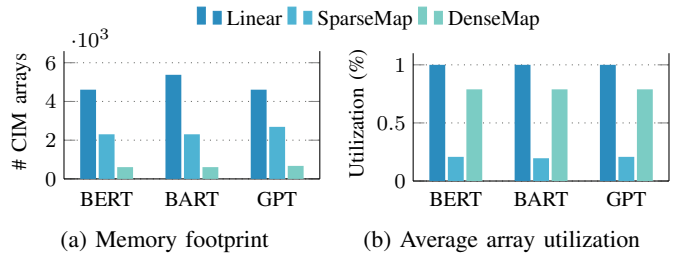
\begin{figure}[bth]
    \centering

    \begin{subfigure}[b]{0.48\columnwidth}
        \centering
        \input{figures/results/reduction}
        \caption{Memory footprint}
        \label{fig:array-count}
    \end{subfigure}
    \hfill
    \begin{subfigure}[b]{0.48\columnwidth}
        \centering
        \input{figures/results/utilization}
        \caption{Average array utilization}
        \label{fig:utilization}
    \end{subfigure}
    \hfill
    \caption{Comparison of memory requirement and resource utilization across different CIM mapping strategies.}
    \label{fig:utilisation_reduction}
\end{figure}

\subsection{Memory Footprint and CIM Arrays Utilization}
\label{subsec:memory-footprint}

In terms of CIM array requirements, Fig.~\ref{fig:array-count} shows that the latency-optimized sparse mapping (\textit{SparseMap}) reduces the number of required arrays by approximately 50\% on average across the three models, compared to the \emph{Linear} baseline. In contrast, the capacity-optimized mapping (\emph{DenseMap}) achieves a significantly higher reduction, requiring 87\% fewer CIM arrays compared to \emph{Linear}, and over 73\% fewer arrays relative to the \textit{SparseMap} configuration.

Fig.~\ref{fig:utilization} compares the array-wise utilization across different configurations. The \emph{Linear} configuration fully utilizes the array capacity. In contrast, the latency-optimized \emph{SparseMap} configuration achieves an average utilization of only 20.4\% relative to \emph{Linear}, with 79.6\% of the array occupied by padded zeros. The capacity-optimized \emph{DenseMap}, on the other hand, significantly improves utilization, reaching an average of 78.8\% of the array capacity. This represents an approximately $3\times$ improvement over \emph{SparseMap}. 
While \emph{DenseMap} substantially improves array utilization, it does not achieve full (i.e., 100\%) utilization due to a mismatch between the Monarch block and CIM array dimensions, i.e., the block size is not an exact multiple of the array dimension. This residual under-utilization can potentially be further reduced by using smaller block sizes to better align with the CIM arrays' dimensions.

\subsection{Performance and Energy Comparison}
\label{subsec:perf-comparison}
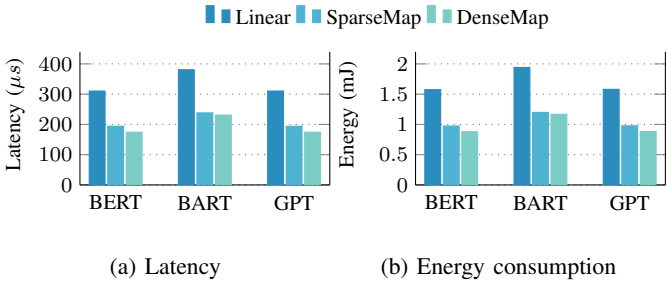
\begin{figure}[bth]
    \centering
    \begin{subfigure}[b]{0.5\columnwidth}
        \input{figures/results/latency}
        \caption{Latency}
        \label{fig:latency}
    \end{subfigure}%
    \hfill
    \begin{subfigure}[b]{0.5\columnwidth}
    \centering
    \input{figures/results/energy}
        \caption{Energy consumption}
       \label{fig:energy}
    \end{subfigure}
    \hfill
    \caption{Latency and energy comparison across configurations.}
    \label{fig:latency-energy}
\end{figure}

Fig.~\ref{fig:latency-energy} compares the latency and energy consumption across all configurations. For the BERT model, the \emph{Linear} configuration achieves a speedup of 16.2$\times$ over the GPU and serves as baseline for the other CIM configurations. Averaged (geomean) across all models, the \emph{SparseMap} configuration improves latency by 1.59$\times$ over \emph{Linear} (see Fig.~\ref{fig:latency}). This improvement stems from three key factors: (i) a reduced number of arithmetic operations due to block-sparse representation, (ii) lower ADC resolution requirements, as only a subset of rows in each array contain non-zero values, and (iii) full parallelism across all CIM arrays, with each block within CIM arrays being processed concurrently (see Sec.~\ref{subsubsec:naive}). 

Although \emph{DenseMap} introduces intra-array sequentiality, its latency is unaffected because (i) all configurations face sequentiality from ADC sharing (assuming one ADC per array; see Fig.~\ref{fig:dse-adc-sharing}) and (ii) it uses lower-resolution ADCs (3b vs. 5b in \emph{SparseMap} and 8b in \emph{Linear}). As a result, latency is reduced by $1.08\times$ over \emph{SparseMap} and $1.73\times$ over \emph{Linear}.

The energy consumption results (Fig.~\ref{fig:energy}) show a similar trend. The baseline \emph{Linear} configuration reduces the energy consumption by three orders of magnitude, compared to the GPU. \emph{SparseMap}, on average, reduces energy consumption by 1.61$\times$ compared to \emph{Linear}, while \emph{DenseMap} achieves a reduction of 1.74$\times$. These gains are primarily attributed to the low-precision ADCs in both configurations.

\subsection{Design Space Exploration}
\label{subsec:dse}
We previously used a fixed system configuration (Table~\ref{tab:params}). Here, we analyze the sensitivity of latency and energy consumption to the number and precision of ADCs/DACs, which account for 60-80\% of the area and energy consumption of MVM in CIM hardware~\cite{negi2022nax}. For the experiments in this section, we use the Accelergy-ADC-plugin~\cite{adc-plugin} to extract the new parameter values and use them with the CIM simulator. 

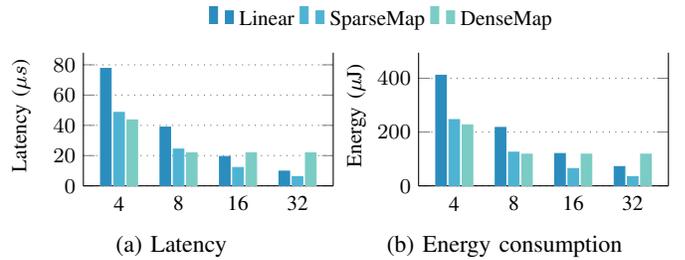
\begin{figure}[t]
    \centering

    \begin{subfigure}[b]{0.5\columnwidth}
        \input{figures/results/latency-adc-sharing}

        \caption{Latency}
        \label{fig:adc-sharing-lat}
    \end{subfigure}%
    \hfill
    \begin{subfigure}[b]{0.5\columnwidth}
    \centering
        \input{figures/results/energy-adc-sharing}
        \caption{Energy consumption}
       \label{fig:adc-sharing-energy}
    \end{subfigure}
    \hfill
    \caption{Latency and energy comparison across varying ADCs per CIM array (X-axis) (BERT model).}
    \label{fig:dse-adc-sharing}
\end{figure}

Fig.~\ref{fig:dse-adc-sharing} compares latency and energy results for the BERT model under different degrees of ADC sharing, varying number of ADCs per array from 4 (one ADC per 64 column) to 32 (one ADC per 8 columns). Our results indicate that sparse configurations, particularly the capacity/density optimized \emph{DenseMap}, exhibit substantial performance improvements over both the \emph{Linear} and \emph{SparseMap} configurations as the ADC sharing degree increases, i.e., fewer ADCs per array (Fig.~\ref{fig:adc-sharing-lat}). This is due to the inherent sequentiality of this configuration, which better aligns with the sequentiality imposed by ADC sharing. Concretely, \emph{DenseMap} achieves a speedup of 1.6$\times$ and 1.1$\times$ over \emph{Linear} and \emph{SparseMap}, respectively, for the 4 ADCs per array system setup. However, as we increase the number of ADCs per array, \emph{Linear} and \emph{SparseMap} record significant gains while the inherent sequentiality of \emph{DenseMap} does not let the latency improve beyond 8 ADCs/array. For instance, for 32 ADCs per array system setup, it performs worse compared to the other configurations. Our latency optimized \emph{SparseMap} achieves the best result, outperforming \emph{DenseMap} by 3.57$\times$ and \emph{Linear} by 1.6$\times$.   

Energy consumption exhibits a similar trend (Fig.~\ref{fig:adc-sharing-energy}). As the number of ADCs is reduced, the energy savings in \emph{DenseMap} become increasingly significant, reinforcing the benefits of our densely packed configuration that exploits temporal sequencing within arrays and minimizes ADC utilization.

\noindent\textbf{ADC/DAC Resolution:} Lowering converter resolution reduces both latency and energy. Reducing the ADC resolution from 8 bits (required by \emph{Linear} for the given array dimensions) to 3 bits (for \emph{DenseMap}) cuts latency and energy by about $2.67\times$.

%% file: figures/results/reduction.tex
\pgfplotsset{compat = newest}
\pgfplotsset{major grid style={dotted,aluminium2!50!black}}
\footnotesize
\begin{tikzpicture}
\begin{axis}
[
    width=0.8\columnwidth,
    height=0.4\textwidth,
    scale only axis,
    ybar=1pt, 
    enlargelimits=0.25,
    enlarge y limits={upper, value=0.1},
    ylabel style={align=center},
    y label style={at={(-0.1,0.53)}},
    ylabel= \# CIM arrays,
    legend style={draw=none, fill=none},
    bar width=6pt,
    legend columns=3,
    ymin=0, ymax=6000,
    ymajorgrids=true,
    xtick style={draw=none},
    grid style=dashed,
    axis x line*=bottom,
    x tick label style={xshift=.0em, yshift=-.4em, rotate=0,anchor=center},
    yminorticks=true,
    legend style={at={(1.2, 1.4)},anchor=north},
    scaled y ticks=base 10:-3,
    symbolic x coords={BERT, BART, GPT},
    xtick=data,
]
\addplot+ [blind_safe_six_scheme_seven_colors_grnblu,draw=none] coordinates {
(BERT, 4608)
(BART, 5376)
(GPT, 4608)
}; 
\addplot+ [blind_safe_five_scheme_seven_colors_grnblu,draw=none] coordinates {
(BERT, 2304)
(BART, 2304)
(GPT, 2688)
}; 
\addplot+ [blind_safe_four_scheme_seven_colors_grnblu, draw=none] coordinates {
(BERT, 608)
(BART, 608)
(GPT, 672)
};  
\legend{Linear, SparseMap, DenseMap}\end{axis}
\end{tikzpicture}

%% file: figures/results/utilization.tex
\pgfplotsset{compat = newest}
\pgfplotsset{major grid style={dotted,aluminium2!50!black}}
\footnotesize
\begin{tikzpicture}
\begin{axis}
[
    width=0.8\columnwidth,
    height=0.4\textwidth,
    scale only axis,
    ybar=1pt, 
    enlargelimits=0.25,
    enlarge y limits={upper, value=0.1},
    ylabel style={align=center},
    y label style={at={(-0.15,0.53)}},
    ylabel= Utilization (\%),
    legend style={draw=none, fill=none},
    bar width=6pt,
    legend columns=5,
    ymin=0, ymax=1,
    ymajorgrids=true,
    grid style=dashed,
    axis x line*=bottom,
    xtick style={draw=none},
    x tick label style={xshift=.0em, yshift=-.3em, rotate=0,anchor=center},
    yminorticks=true,
    legend style={at={(0.3, 1.4)},anchor=north},
    symbolic x coords={BERT, BART, GPT},
    xtick=data,
]
\addplot+ [blind_safe_six_scheme_seven_colors_grnblu,draw=none] coordinates {
(BERT, 1)
(BART, 1)
(GPT, 1)
}; 
\addplot+ [blind_safe_five_scheme_seven_colors_grnblu,draw=none] coordinates {
(BERT, 0.208)
(BART, 0.196)
(GPT, 0.208)
}; 
\addplot+ [blind_safe_four_scheme_seven_colors_grnblu, draw=none] coordinates {
(BERT, 0.789)
(BART, 0.789)
(GPT, 0.789)
}; 

\end{axis}
\end{tikzpicture}

%% file: figures/results/latency.tex
\pgfplotsset{compat = newest}
\pgfplotsset{major grid style={dotted,aluminium2!50!black}}
\footnotesize
\begin{tikzpicture}
\begin{axis}
[
    width=0.75\columnwidth,
    height=0.4\textwidth,
    scale only axis,
    ybar=1pt,
    enlargelimits=0.2,
    enlarge y limits={upper, value=0.1},
    ylabel style={align=center},
    y label style={at={(-0.19,0.53)}},
    ylabel= Latency ($\mu s$),
    legend style={draw=none, fill=none},
    bar width=6pt,
    legend columns=5,
    ymin=0, ymax=400,
    ymajorgrids=true,
    grid style=dashed,
    axis x line*=bottom,
    xtick style={draw=none},
    x tick label style={xshift=.0em, yshift=-.3em, rotate=0,anchor=center},
    yminorticks=true,
    legend style={at={(1.2, 1.4)},anchor=north},
    symbolic x coords={BERT, BART, GPT},
    xtick=data,
]
\addplot+ [blind_safe_six_scheme_seven_colors_grnblu] coordinates {
(BERT, 310.400)
(BART, 380.800)
(GPT, 310.400)
}; 
\addplot+ [blind_safe_five_scheme_seven_colors_grnblu] coordinates {
(BERT, 194.145)
(BART, 238.178)
(GPT, 194.145)
}; 
\addplot+ [blind_safe_four_scheme_seven_colors_grnblu] coordinates {
(BERT, 174.362)
(BART, 230.880)
(GPT, 174.362)
};  
\legend{Linear, SparseMap, DenseMap}
\end{axis}
\end{tikzpicture}

%% file: figures/results/energy.tex
\pgfplotsset{compat = newest}
\pgfplotsset{major grid style={dotted,aluminium2!50!black}}
\footnotesize
\begin{tikzpicture}
\begin{axis}
[
     width=0.75\columnwidth,
    height=0.4\textwidth,
    scale only axis,
    ybar=1pt, 
    enlargelimits=0.2,
    enlarge y limits={upper, value=0.1},
    ylabel style={align=center},
    y label style={at={(-0.2,0.53)}},
    ylabel= Energy (mJ),
    legend style={draw=none, fill=none},
    bar width=6pt,
    legend columns=5,
    ymin=0, ymax=2,
    ymajorgrids=true,
    grid style=dashed,
    axis x line*=bottom,
    xtick style={draw=none},
    x tick label style={xshift=.0em, yshift=-.3em, rotate=0,anchor=center},
    yminorticks=true,
    legend style={at={(1.2, 1.4)},anchor=north},
    symbolic x coords={BERT, BART, GPT},
    xtick=data,
]
\addplot+ [blind_safe_six_scheme_seven_colors_grnblu] coordinates {
    (BERT, 1.57508)
    (BART, 1.94144)
    (GPT,  1.57984)
}; 

\addplot+ [blind_safe_five_scheme_seven_colors_grnblu] coordinates {
    (BERT,  0.9743525)
    (BART, 1.1989925)
    (GPT,   0.9773275)
}; 

\addplot+ [blind_safe_four_scheme_seven_colors_grnblu] coordinates {
    (BERT,  0.8804675)
    (BART, 1.16844)
    (GPT,   0.8822525)
}; 
\end{axis}
\end{tikzpicture}

%% file: figures/results/latency-adc-sharing.tex
\pgfplotsset{compat = newest}
\pgfplotsset{major grid style={dotted,aluminium2!50!black}}
\footnotesize
\begin{tikzpicture}
\begin{axis}
[
    width=0.75\columnwidth,
    height=0.4\textwidth,
    scale only axis,
    ybar=1pt, 
    enlargelimits=0.2,
    enlarge y limits={upper, value=0.1},
    ylabel style={align=center},
    y label style={at={(-0.18,0.53)}},
    ylabel= Latency ($\mu s$),
    legend style={draw=none, fill=none},
    bar width=4pt,
    legend columns=5,
    ymin=0, ymax=80,
    ymajorgrids=true,
    grid style=dashed,
    axis x line*=bottom,
    xtick style={draw=none},
    x tick label style={xshift=.0em, yshift=-.3em, 
    rotate=0,anchor=center},
    yminorticks=true,
    legend style={at={(1.2, 1.4)},anchor=north},
    symbolic x coords={4, 8, 16, 32},
    xtick=data,
]
\addplot+ [blind_safe_six_scheme_seven_colors_grnblu] coordinates {
(4, 77.600)
(8, 38.800)
(16, 19.400)
(32, 9.700)
}; 
\addplot+ [blind_safe_five_scheme_seven_colors_grnblu] coordinates {
(4, 48.536)
(8, 24.268)
(16, 12.134)
(32, 6.067)
}; 
\addplot+ [blind_safe_four_scheme_seven_colors_grnblu] coordinates {
(4, 43.590)
(8, 21.795)
(16, 21.795)
(32, 21.795)
};  
\legend{Linear, SparseMap, DenseMap}
\end{axis}
\end{tikzpicture}

%% file: figures/results/energy-adc-sharing.tex
\pgfplotsset{compat = newest}
\pgfplotsset{major grid style={dotted,aluminium2!50!black}}
\footnotesize
\begin{tikzpicture}
\begin{axis}
[
     width=0.75\columnwidth,
    height=0.4\textwidth,
    scale only axis,
    ybar=1pt, 
    enlargelimits=0.2,
    enlarge y limits={upper, value=0.1},
    ylabel style={align=center},
    y label style={at={(-0.18,0.53)}},
    ylabel= Energy ($\mu$J),
    legend style={draw=none, fill=none},
    bar width=4pt,
    legend columns=5,
    ymin=0, ymax=450,
    ymajorgrids=true,
    grid style=dashed,
    axis x line*=bottom,
    xtick style={draw=none},
    x tick label style={xshift=.0em, yshift=-.3em, rotate=0,anchor=center},
    yminorticks=true,
    legend style={at={(1.2, 1.4)},anchor=north},
    symbolic x coords={4, 8, 16, 32},
    xtick=data,
]
\addplot+ [blind_safe_six_scheme_seven_colors_grnblu] coordinates {
(4, 411.080)
(8, 217.080)
(16, 120.080)
(32, 71.580)
}; 
\addplot+ [blind_safe_five_scheme_seven_colors_grnblu]  coordinates {
(4, 246.306)
(8, 124.965)
(16, 64.295)
(32, 33.960)
}; 
\addplot+ [blind_safe_four_scheme_seven_colors_grnblu]  coordinates {
(4, 226.608)
(8, 117.631)
(16, 117.631)
(32, 117.631)
}; 

\end{axis}
\end{tikzpicture}

%% file: content/related-work.tex
\section{Related Work}
\label{sec:relwork}

\noindent 
\textbf{Structured matrices in digital accelerators}: \cite{fan2022adaptable} proposes a HW-friendly model leveraging butterfly sparsity in attention and FFN layers, along with an adaptable, unified butterfly engine. However, their design targets only digital ASICs and is not compatible with analog CIM.

\noindent 
\textbf{Sparse attention accelerators in CIM:} ASADI~\cite{li2024asadi} proposes a diagonal compression format and a sparse attention accelerator within the CIM paradigm. TranCIM~\cite{tu2022trancim} also targets block-sparse attention in CIM and proposes a scheduler that reduces attention's complexity. 
While ASADI and TranCIM support block-diagonal sparsity, they overlook the permutations and access patterns from densely packing such structures onto CIM arrays. Moreover, they target only sparse attention with dynamic data sparsity and no stationary weights, which makes their proposals orthogonal to ours.. 
In contrast, we are the first to address the mapping and scheduling of parameterized block-diagonal sparsity in both attention and FFN layers for CIM.

%% file: content/conclusions.tex
\section{Conclusions}
\label{sec:conclusions}
We present an automated framework for efficiently mapping structured sparse transformer models on analog CIM accelerators. Starting from a dense model, we apply dense-to-sparse transformations to produce Monarch matrices, which are then mapped to CIM crossbars using latency- and capacity-optimized strategies. A mapping-aware scheduler orchestrates execution by generating addresses and issuing CIM commands. Our evaluation of three transformer models shows up to $1.74\times$ speedup and energy reduction over their baseline dense models. While our simulation framework does not directly support area modeling, the observed reduction in the number of CIM arrays (Fig.~\ref{fig:array-count}) and ADCs (Fig.~\ref{fig:dse-adc-sharing}) serve as a reliable proxy, suggesting over 4$\times$ area savings. 

\section*{Acknowledgment}
The authors acknowledge the financial support by the Federal Ministry of Research, Technology and Space of Germany and by Sächsische Staatsministerium für Wissenschaft, Kultur und Tourismus in the programme Center of Excellence for AI-research „Center for Scalable Data Analytics and Artificial Intelligence Dresden/Leipzig", project identification number: ScaDS.AI, and funded by the German Research Council (DFG) through the HetCIM project (502388442), and the CRC/TRR 404-Active 3D (528378584).
